\documentclass[12pt,a4paper]{article}
\newlength{\HFPP}       \HFPP5.4mm
\makeatletter
\@addtoreset{equation}{section}
\makeatother

\def\lam{\frac{2\Lambda}{\pi}}

\def\a{\alpha}
\def\al{\widehat{\alpha}}
\def\n{\frac{\al}{2\pi}}
\def\x{\widehat{x}}

\def\g{\gamma}

\def\o{\omega(t)}
\def\dv{|0)}

\def\nddv{{(\tilde 0|}}
\def\La{\Lambda}
\def\be{\begin{equation}}
\def\ee{\end{equation}}
\def\bea{\begin{eqnarray}}
\def\eea{\end{eqnarray}}
\def\r#1{(\ref{#1})}

\def\nn{\nonumber\\}
\def\l{\lambda}

\def\m{\mu}
\def\emph#1{{\sl #1}}
\def\textbf#1{{\bf #1}}

\def\emph#1{{\sl #1}}
\def\PRL#1#2#3{{\sl Phys. Rev. Lett.} {\bf#1} (#2) #3}
\def\JMP#1#2#3{{\sl J. Math. Phys.} {\bf #1} (#2) #3}
\begin{document}

\begin{titlepage}

\def\thefootnote{\fnsymbol{footnote}}
\setlength{\baselineskip}{18pt}
\begin{flushleft}
OUTP-96-16S \hfill solv-int/9604005\\
ITP-UH-06/96
\end{flushleft}

\vspace*{\fill}

\begin{center}
{\large\bf Painlev\'e Transcendent Describes Quantum Correlation\\[.5cm]
Function of the XXZ Antiferromagnet away from the\\[.5cm]
free-fermion point}
\vfill

{\sc F.H.L.\,E\ss{}ler$^1$, H. Frahm$^2$, A.R. Its$^3$ and V.E.
Korepin$^{4-6}$}\\[2em]

$^1${\sl Department of Physics, Theoretical Physics, 1 Keble Road,\\
	Oxford OX1 3NP, Great Britain}\\[8pt]

$^2${\sl Institut f\"ur Theoretische Physik, Universit\"at Hannover,\\
	D-30167~Hannover, Germany}\\[8pt]

$^3${\sl Department of Mathematical Sciences,\\
	Indiana University-Purdue University at Indianapolis (IUPUI),\\
	Indianapolis, IN 46202--3216, U. S. A.}\\[8pt]

$^4${\sl Institute for Theoretical Physics, State University of New
         York at Stony Brook,\\
         Stony Brook, NY 11794--3840, U. S. A.}\\
$^5${\sl Yukawa Institute for Theoretical Physics, Kyoto
University, Kyoto 606, Japan}\\
$^6${\sl Sankt Petersburg Department of Mathematical Institute of
Academy of Sciences of Russia}

\vfill
ABSTRACT
\end{center}

\begin{quote}
We consider quantum correlation functions of the antiferromagnetic
spin-$\frac{1}{2}$ Heisenberg XXZ spin chain in a magnetic field. We
show that for a magnetic field close to the critical field $h_c$ (for
the critical magnetic field the ground state is ferromagnetic) certain
correlation functions can be expressed in terms of the solution of the
Painlev\'e V transcendent. This establishes a relation between
solutions of Painlev\'e differential equations and quantum correlation
functions in models of {\sl interacting} fermions. Painlev\'e
transcendents were known to describe correlation functions in models
with free fermionic spectra.

\end{quote}

\vfill
PACS-numbers:
75.10.Jm 
02.30.Jr 

\setcounter{footnote}{0}
\end{titlepage}

\section{Introduction}

In this letter we continue our investigation of zero-temperature
correlation functions of the XXZ Heisenberg model in the
critical regime $-1<\Delta<1$ in an external magnetic field.
The XXZ hamiltonian is given by
\begin{equation}
   {\cal H} = \sum_{j} \sigma_j^x \sigma_{j+1}^x
                     + \sigma_j^y \sigma_{j+1}^y
                     + \Delta\ (\sigma_j^z \sigma_{j+1}^z-1)
                     - h \sigma_j^z\ ,
   \label{xxz}
\end{equation}
where the sum is over all integers $j$, $L$ is the length of the
lattice, $\sigma^\alpha$ are Pauli matrices
and $h$ is an external magnetic field. For later convenience we define
$\Delta = \cos(2\eta)$, where ${\pi\over 2}<\eta< \pi$. The free
fermionic point in this notation is $\eta=\frac{3\pi}{4}$. The model
\r{xxz} can be solved by means of the Bethe Ansatz, which yields a
description of the spectrum and eigenstates (with $N$ down spins and
$L-N$ up spins) in terms of the roots of the following set of coupled
algebraic equations \cite{orbach,yaya}
\be
\left(\frac{\sinh(\l_j-i\eta)}{\sinh(\l_j+i\eta)}\right)^L=-\prod_{k=1}^N
\frac{\sinh(\l_k-\l_j+2i\eta)}{\sinh(\l_k-\l_j-2i\eta)}\ ,
j=1\ldots N\ .
\label{bae}
\ee
For the case $\Delta > -1$ it was proved by C.N. Yang and C.P. Yang in
\cite{yaya} that the ground state is characterized by a set of
{\sl real} $\lambda_j$ subject to the equations \r{bae}.
In the thermodynamic limit the ground state is described by means of an
integral equation for the density of spectral parameters $\rho(\lambda)$
\bea
2\pi \rho(\lambda) - \int_{-\Lambda}^{\Lambda} d\mu\
K(\lambda ,\mu)\ \rho(\mu)\ &=&
\frac{-\sin(2\eta)}{\sinh(\lambda -i\eta) \sinh(\lambda +i\eta)}\nn
K (\m, \l ) &=& \frac{\sin(4\eta)}{\sinh(\mu- \lambda
+2i\eta)\sinh(\mu - \lambda -2i\eta)}\ .
\label{gsie}
\eea
Here the integration boundary $\La$ is a function of the magnetic
field $h$. The physical picture of the ground state is that of a
filled Fermi sea with boundaries $\pm \Lambda$. The dressed energy of
a particle in the sea is given by the solution of the integral
equation
\be
\epsilon(\lambda) - {1\over 2\pi}\int_{-\Lambda}^{\Lambda} d\mu\
K(\lambda ,\mu)\ \epsilon(\mu)\ = 2h -{2 (\sin(2\eta))^2\over
\sinh(\lambda -i\eta) \sinh(\lambda +i\eta) }\ .
\ee
The condition that the dressed energy vanishes at the Fermi edge
$\epsilon(\pm\La) = 0$ determines $\La$ as a function of the magnetic
field $h$.
It was shown in \cite{yaya} that the ground state of \r{xxz}
for $|\Delta|<1$ is partially magnetized for magnetic fields
$h<h_c=4\cos^2(\eta)$. For larger magnetic fields $h>h_c$ the ground
state is the saturated ferromagnetic state. For $h\to 0$ the integration
boundary $\La$ tends to $\infty$, whereas for $h\to h_c$ $\La\to 0$.
Below we only consider the region $h<h_c$ and in particular the
limiting case $h\rightarrow h_c$.

The subject of this letter is the generating functional of correlation
functions
\begin{equation}
G(m)=\langle \exp(\a Q_1(m))\rangle = \langle 0| \exp(\a\sum_{j=1}^m
\sigma_j^-\sigma_j^+)|0\rangle= \langle 0| \exp(\a\sum_{j=1}^m
\frac{1-\sigma_j^z}{2})|0\rangle\ ,
\label{gm}
\end{equation}
where $|0\rangle$ is the ground state.

Various correlation functions can be obtained from $\langle \exp(\a
Q_1(m))\rangle$, ({\sl e.g.}, the Ferromagnetic String Formation
Probability \cite{efik:95a} which corresponds to setting
$\a=-\infty$) or
\be
\langle 0|\sigma_m^z\sigma_1^z|0\rangle = 2{\widehat\Delta}\langle
0|\frac{\partial^2}{\partial\a^2}\bigg|_{\a=0} \exp(\a
Q_1(m))|0\rangle + 1-4\int_{-\Lambda}^\Lambda d\lambda\ \rho(\lambda)\
,
\label{szsz}
\ee
where $\widehat{\Delta}$ is the lattice laplacian acting on a function
$f(j)$ defined on the lattice as
$\widehat{\Delta}f(j)=f(j)+f(j-2)-2f(j-1)$ and $\rho(\l)$ is defined
in \r{gsie}. In what
follows we will consider $G(m)$ in the limit $h\to h_c$. As is shown
below $G(m)$ is connected to the solution of a Painlev\'e V
transcendent. The connection of Painlev\'e transcendents and
integrable models with free-fermionic spectra is well established
\cite{pain,ff2,barry}. However, we want to emphasize that in the
present case we are dealing with a theory of {\sl interacting}
fermions, so that the connection is novel.

\section{Determinant Representation}

In a recent paper \cite{efik:95} we used the approach invented in
\cite{kore:87,itsx:90} (for a detailed exposition of this method see
\cite{vladb}) to derive the following representation of
$G(m)$ in terms of the determinant of a Fredholm integral
operator.
\be
\langle 0|\exp(\a Q_1(m))|0\rangle
=\frac{\nddv\det{\left(1+\widehat{V}\right)}\dv}
{\det{\left(1-\frac{1}{2\pi}\widehat{K}\right)}}\ .
\label{final}
\ee
Here $\widehat{K}$ and $\widehat{V}$ are Fredholm integral operators
with kernels given by \r{gsie} and
\bea
   &&V (\l, \m)  =  -\frac{\sin(2\eta)}{2\pi\sinh(\l-\m)}
        \bigg\{ {1\over\sinh(\l-\m +2i\eta)}
         + {e_2^{-1} (\l) e_2(\m) \over\sinh(\l - \m -2i\eta)} \nn
        &&\qquad + \exp(\a +\varphi_4(\mu) - \varphi_3(\lambda))
        \bigg({1\over\sinh(\l-\m -2i\eta)}
         + {e_1^{-1} (\mu) e_1(\lambda)\over\sinh(\lambda-\mu+2i\eta)}
        \bigg)
        \bigg\}\ ,
\label{vcont}
\eea
where
\be
  e_2 (\l) =
        \bigg( {\sinh(\l + i\eta) \over \sinh(\l - i\eta)}
        \bigg)^{m} e^{\varphi_2 (\l )}\ ,
  \quad
  e_1 (\l) =
        \bigg( {\sinh(\l - i\eta) \over \sinh(\l + i\eta)}
        \bigg)^{m} e^{\varphi_1 (\l )}\ .
\ee
The quantities $\varphi_j(\l)$ are bosonic quantum fields
\cite{kore:87} (they are called ``dual quantum fields'')
defined by
\bea
   \varphi_a(\lambda) &=& p_a(\lambda) + q_a(\lambda)\ ,\
   \nddv q_a(\l) = 0 = p_a(\l) \dv \ , \nddv 0)=1\ ,\ a=1\ldots 4\ ,
   \nonumber \\[8pt]
   {[q_b(\mu), p_a(\lambda)]} &=&\!\!\!\!\left(
        \matrix{1&0&1&0\cr 0&1&0&1\cr 0&1&1&1\cr 1&0&1&1\cr}
   \right)_{ab} \ln(h(\lambda ,\mu)) + \left(
        \matrix{1&0&0&1\cr 0&1&1&0\cr 1&0&1&1\cr 0&1&1&1\cr}
   \right)_{ab} \ln(h(\mu, \lambda)),
\label{dfcom}
\eea
where $h(\l,\m)= {\sinh(\l-\m+2i\eta)\over i\sin(2\eta)}$. Note that
the dual fields have the important property that they commute
\be
[\varphi_a(\l),\varphi_b(\mu)]=0\ .
\ee

\section{The limit of strong magnetic field}

For large magnetic fields $h\rightarrow h_c, h<h_c$ the integration boundary
$\La$ tends to zero according to
\be
\La=\frac{1}{2}|\tan(\eta)|\sqrt{h_c-h}+O(h_c-h).
\ee
This fact can be used to essentially simplify the above representation
as was first done for a different correlation function in
\cite{efik:95a}.
We first expand the kernel $V(\l,\m)$ for small $|\l|,|\mu|\leq\La$.
We define $y=m|\cot(\eta)|$ as a shorthand notation. We then observe the
following simplification of the commutation relations between the
``momenta'' $p_a(\l)$ and ``coordinates'' $q_a(\l)$
\be
   {[q_b(\mu), p_a(\lambda)]} =\left(
        \matrix{0&0&1&-1\cr 0&0&-1&1\cr -1&1&0&0\cr 1&-1&0&0\cr}
   \right)_{ab} i\cot(2\eta) (\m-\l)+{\cal O}(\La^2)\ .
\label{dfcom2}
\ee
This allows us to reduce the number of dual fields from four to two
{\sl via} the identification $\varphi_2(\l)=-\varphi_1(\l)$ and
$\varphi_4(\l)=-\varphi_3(\l)$. Furthermore the r.h.s. of the
commutators \r{dfcom2} is a linear function of $\l-\mu$ so that we can
choose a representation such that $\varphi_a(\l)$ are linear functions in
$\l$
\be
\varphi_a(\l)=\varphi_a+\varphi_a' \l\ ,\quad
p_a(\l)=p_a+p_a' \l\ ,\quad
q_a(\l)=q_a+q_a' \l\ ,\ a=1,3
\ee
where $[p_a,q_b]=0$, $[p_a',q_b']=0$, and
\be
[q_1',p_3]=-i\cot(2\eta)=-[q_3',p_1]\ ,\quad
[q_3,p_1']=-i\cot(2\eta)=-[q_1,p_3']\ .
\label{qm}
\ee
As $\varphi_1(\l)$ appears in \r{vcont} only in the combination
$\varphi_1(\l)-\varphi_1(\mu)$ the quantity $\varphi_1$ drops out.
This in turn implies that $p_3'$ and $q_3'$ commute with all remaining
operators and thus will not contribute to the expectation value w.r.t.
$\nddv$ and $\dv$. Therefore we can drop them everywhere.
In the next step we perform a similarity transformation with
$\exp(i\l (y+\frac{i}{2}\varphi_1'))$, which leaves
the determinant of the Fredholm integral operator invariant but brings
the kernel to a more symmetric form, in which the dual fields now only
enter {\sl via} the expressions $\al:=\a-2\varphi_3$ and
$\x:=y+\frac{i}{2}\varphi_1'$. In what follows it is crucial that the
quantum operators in the dual bosonic Fock space $\x$ and $\al$ are
commuting objects and no problems with operator orderings occur before
we evaluate the expectation value w.r.t. $\nddv$ and $\dv$.

Putting now everything together we arrive at the following simplified
representation valid in the limit $h\rightarrow h_c$
\be
\langle\exp(\a Q_1(m))\rangle
={\nddv\det{\left(1+\widehat{V_0}\right)}\dv}
{\left(1+\frac{2\La}{\pi}\cot(2\eta)+{\cal O}(\La^2)\right)}\ ,
\label{simpl}
\ee
where the kernel of $\widehat{V_0}$ is given by
\be
V_0(\l,\m)=\left(\exp(\al)-1\right)
\frac{\sin((\l-\mu)\x)}{\pi(\l-\mu)}\ .
\ee
Here we use the following notations for the dual fields
\bea
\al &=& \a +\al_q+\al_p\ ,\quad \x=y +\x_q+\x_p\ ,
\nddv \x_q=\nddv \al_q=0\ ,\quad \x_p\dv=\al_p\dv=0\ ,\nn
\lbrack\x_q},{\al_p\rbrack &=& \lbrack\al_q,\x_p\rbrack=-\cot(2\eta)\
,\ [\al ,\x] = 0\ .
\label{xa}
\eea
All other commutators vanish. We would like to emphasize that $\al$
and $\x$ commute, which is important for the further analysis.

\section{Connection with Painlev\'e V}
Let us define a new variable $t=\Lambda \x$ and consider the object
\be
\sigma_0(t)=t\frac{d}{dt}\ln(\det(1+\widehat{V}_0)).
\ee

In \cite{ff2} it is shown that $\sigma_0$ obeys a
Painlev\'e V differential equation in the case where $\alpha$ and $x$
are real numbers. In our case $\al$ and $\x$ are quantum operators,
but due to the fact that they are commuting we still can follow through
the derivation of \cite{ff2}. We thus find that $\sigma_0$ obeys the
following nonlinear differential equation
\be
\left(t\frac{d^2\sigma_0(t)}{dt^2}\right)^2=
-4\left(t\frac{d\sigma_0(t)}{dt}-\sigma_0(t)\right)
\left(4t\frac{d\sigma_0(t)}{dt}+ (\frac{d\sigma_0(t)}{dt})^2
-4\sigma_0(t)\right)\ ,
\label{pains}
\ee
which is identified (see \cite{ff2})  as a $\tau$ - function
form of the Painlev\'e V equation.
Rewriting \r{pains} in terms of the function $y_0(t)$ defined through
\be
\sigma_0(t)=-4itu(t)+\frac{u^2(t)}{y_0(t)}(y_0(t)-1)^2\ ,\quad
u(t)=\frac{4ity_0(t)-t\frac{dy_0(t)}{dt}}{2(y_0(t)-1)^2}\
\ee
one obtains (see again \cite{ff2}) standard form of the Painlev\'e V
differential equation for the function $\omega(t)=y_0(\frac{t}{2})$
\be
\frac{d^2\omega(t)}{dt^2}=\left(\frac{d\omega(t)}{dt}\right)^2
\frac{3\o-1}{2\o(\o-1)}+\frac{2\o(\o+1)}{\o-1}+\frac{2i\o}{t}-
\frac{1}{t}\frac{d\o}{dt}\ .
\label{p5}
\ee
The large $t$ asymptotics of the solution of the above equations are
known (see \cite{barry,wid,bb,sul}) and can be used to extract the
large distance asymptotics of our Fredholm determinant. Combining the
results of \cite{barry} (asymptotics for $\sigma_0$) and \cite{wid,bb}
(constant term) we obtain that
\bea
\ln(\det(1+\widehat{V}_0)) &=&
\frac{2}{\pi}\La\x\al+\frac{\al^2}{2\pi^2}\ln(4\La\x)+2\ln(g(\n))
+\frac{\al^3}{8\pi^3\x\La}\nn
&&-\frac{1}{\x^2}\left[\frac{\al^2}{32\pi^2\La^2}\cos(4\theta)
+\frac{5\al^4}{128\pi^4\La^2}\right]+{\cal O}(\frac{1}{{\La^3}y^3})\ ,
\label{res}
\eea
where the distance $y$ must be large (the product of $y$ and the small
parameter $\Lambda$ should go to infinity) and where
\bea
g(\nu)&=&e^{(1+\g)\nu^2}\prod_{n=1}^\infty(1+\frac{\nu^2}{n^2})^n
e^{-\frac{\nu^2}{n}} \ ,\nn
  &=& \exp\left(\nu^2
  - {1\over2} \int_0^{\nu^2} dt\ \left[
        \psi(1+i\sqrt{t}) + \psi(1-i\sqrt{t}) \right]\right)\ , \nn
4\theta&=&4\La\x+\frac{2}{\pi}\al\ln(4\La\x)
-4{\rm arg}\Gamma(\frac{i\al}{2\pi})\ ,
\label{theta}
\eea
where $\psi(z)={d\over dz} \ln\Gamma(z)$ is the digamma function and $\g$
is Euler's constant. Formula \r{theta} for the phase $\theta$ of the
corresponding
solution of the Painlev\'e V equation \r{p5} was also obtained in \cite{sul}.
In order to obtain the large-distance asymptotics of the correlation
function we still have to evaluate the expectation value in the dual
bosonic Fock space. It is easiest to evaluate the quantities $\langle
Q_1(m)^k\rangle = \langle\frac{\partial^k}{\partial\a^k}\bigg|_{\a=0}
e^{\a Q_1(m)}\rangle$ directly. Expanding
\bea
e^{\frac{\al^2}{2\pi^2}\ln(4\La \x)} &=& e^{\frac{\al^2}{2\pi^2}\ln(4\La
y)} \left[1+\frac{\al^2}{2\pi^2}\frac{\x_p+\x_q}{y}
+{\cal O}(y^{-2})
\right]\nn
e^{\frac{\al^3}{8\pi^3\La \x}}&=&1+ \frac{\al^3}{8\pi^3\La y} + {\cal
O}({\La^{-1}}y^{-2})\  ,
\label{explog}
\eea
we obtain the leading terms in the asymptotic decompositions as is
shown in the Appendix
\bea
\langle Q_1(m)\rangle\bigg|_{\rm lead} &=&\lam m|\cot\eta|
{\left(1+\frac{6\La}{\pi}\cot(2\eta)+{\cal O}(\La^2)\right)}\nn
\langle (Q_1(m))^2\rangle\bigg|_{\rm lead} &=&
\left[(\lam m|\cot\eta| )^2 +\frac{\ln(4\La m|\cot\eta|)}{\pi^2}\right]
(1+\frac{8\La\cot(2\eta)}{\pi}+{\cal O}(\La^2))\nn
&&+(1+\gamma)(\frac{1}{\pi^2}+\frac{8\La\cot(2\eta)}{\pi^3})
+\frac{6\La\cot(2\eta)}{\pi^3}+{\cal O}(\La^2)\ .
\label{qs}
\eea
Analogous formulas can be derived for $(Q_1(m))^k$ for $k>2$.
Using \r{szsz} we are now in a position to determine the
large-distance asymptotics of the $\langle\sigma^z_m\sigma^z_1\rangle$
correlation functions. By acting with (twice) the lattice laplacian on
\r{qs} we obtain 
\be
-\frac{2}{m^2\pi^2}\left(1+\frac{8\La\cot(2\eta)}{\pi}+{\cal
O}(\La^2)\right).
\ee
This does not yet include the contribution from the $\cos$-term in
\r{res}, which is very small as far as $\langle (Q_1(m))^2\rangle$ is
concerned but becomes important upon differentiation. The leading
contribution can be obtained by the methods explained in the appendix
and is found to be
\be
\left(\frac{2}{\pi^2}+{\cal O}(\La)\right)\cos\left(4\La m|\cot\eta|(
1+{\cal O}(\La))\right)\frac{1}{m^\theta}\ ,
\ee
where
\be
\theta = 2+\frac{8\La\cot(2\eta)}{\pi}+{\cal O}(\La^2).
\ee
Our final result for the first three terms of the asymptotics of the 
correlation function is thus
\bea
\langle\sigma^z_m\sigma^z_1\rangle &=&
1-\frac{8\La|\cot(\eta)|}{\pi}
-\frac{2}{m^2\pi^2}\left(1+\frac{8\La\cot(2\eta)}{\pi}\right)
+\frac{2}{\pi^2m^\theta}\cos(4\La m|\cot\eta|)+\ldots\ ,\nn
\eea
where the errors in the $\La$-expansion are given above.
This agrees with the result obtained by means of finite-size
corrections and Conformal Field Theory. 

\section{Summary and Conclusion}

In this letter we have established a connection between the generating
functional of correlation functions $G(m)$ \r{gm} (and thus the
correlator $\langle \sigma^z_m\sigma^z_1\rangle$) of the spin
$\frac{1}{2}$ Heisenberg XXZ model in a magnetic field close to
critical and the Painlev\'e V differential equation. Painlev\'e
transcendents were known to describe correlation functions for models
with free-fermionic spectra \cite{pain,barry}. For generic magnetic fields
in the XXZ spin chain the general approach of \cite{vladb} should be
followed: the determinant representation \r{final}, \r{vcont} should
be used to embed the quantum correlation function in an integrable
system of integro-difference equations and one then should solve the
associated Riemann-Hilbert problem.

\section*{Acknowledgements}
We are grateful to Barry McCoy for helpful discussions and to Gordon
Chalmers for consultations on normal ordering.
This work was partially supported by the Deutsche
For\-schungs\-gemein\-schaft under Grant No.\ Fr~737/2--1 and by the
National Science Foundation (NSF) under Grants No.\ PHY-9321165 and
No.\ DMS-9501559.
F.H.L.E.\ is supported by the EU under Human Capital and Mobility
fellowship grant ERBCHBGCT940709.

\appendix
\section{Appendix}

In this appendix we discuss how to evaluate the expectation value
with respect to the dual quantum fields and explicitly evaluate the
quantities $\langle(Q_1(m))^k\rangle$. According to \r{simpl}
\be
\langle (Q_1(m))^k\rangle = 
\nddv\frac{\partial^k}{\partial\a^k}\bigg|_{\a=0}
\det{\left(1+\widehat{V_0}\right)}\dv
{\left(1+\frac{2\La}{\pi}\cot(2\eta)+{\cal O}(\La^2)\right)}\ .
\label{app1}
\ee
Using \r{res} for the asymptotics of the logarithm of the determinant
and expanding according to \r{explog} we see that in order to get the
leading asymptotics we need to evaluate the derivatives of expectation
values of the form
\be
\nddv e^{\frac{2}{\pi}\La\x\al}e^{\frac{\al^2}{2\pi^2}\ln(4\La y)}
(g(\n))^2\left[1+\frac{\al^2}{2\pi^2}\frac{\x_p+\x_q}{y}\right]
\left[1+ \frac{\al^3}{8\pi^3\La y}\right]\dv\ .
\label{app2}
\ee
Using the commutation relations \r{xa} we see that we can bring all
terms into the form (we use that $\nddv \x_q =0$)
\be
\nddv \x_p^m e^{\frac{2}{\pi}\La \x\al} F(\al)\dv\ ,
\label{appp}
\ee
where $F(\al)$ is only a function of $\al$ and contains no $\x_p$'s or
$\x_q$'s. The central identities we will use in order to evaluate the
expectation values are
\bea
z_m=\nddv \x_p^me^{\frac{2}{\pi}\La \x\al} F(\al)\dv &=&
\frac{1}{\kappa}\nddv 
\left(\frac{\x_p}{\kappa}+\frac{2\La\cot(2\eta)}{\pi\kappa}y\right)^m
e^{\frac{2\La\a}{\pi\kappa}(y+\x_p)}F(\al)\dv.\nn
\label{a1}
\eea
where $\kappa = {1-\lam\cot(2\eta)}$.
The identities are established via induction. The induction start
$m=0$ is proved as follows. Expanding the exponential and using that
$\nddv \x_q =0$, $\al_p\dv=0$ and $[\al_p,\al_q]=0$ (to move all
$\al_p$'s to the right) we obtain
\be
z_0=\nddv \sum_{n=0}^\infty\frac{1}{n!}\left(\lam\right)^n
(y+\x_p)^n (\a+\al_q)^n F(\al)\dv\ .
\ee
Expanding
\be
\nddv (y+\x_p)^n (\a+\al_q)^n = \sum_{k=0}^n
\frac{n!}{k!(n-k)!} \nddv (y+\x_p)^n \al_q^k \a^{n-k}
\ee
and then using the commutation relations
\be
\nddv [f(\x_p),\al_q^k] =\nddv  (\cot(2\eta))^k f^{(k)}(\x_p)\ ,
\ee
where $f^{(k)}$ is the k'th derivative of the function $f$,
we arrive at
\bea
z_0&=&\sum_{n=0}^\infty\sum_{k=0}^n \left(\lam\right)^n
\frac{n!(\cot(2\eta))^k\a^{n-k}}{k![(n-k)!]^2} \nddv (y+\x_p)^{n-k}
F(\al)\dv\nn
\eea
We now use the integral representation
\be
\frac{1}{(n-k)!} = \frac{1}{2\pi i}\oint dt\ \frac{e^t}{t^{n-k+1}}\ ,
\ee
where the integration contour is a small circle around the origin (and
we integrate in the mathematically positive direction) in order to be
able to perform the $k$-summation (which is of the form of a
binomial sum)
\bea
z_0&=&\frac{1}{2\pi i}\oint dt \frac{e^t}{t}
\sum_{n=0}^\infty\left(\lam\right)^n
\nddv [\cot(2\eta)+\frac{\a}{t}(y+\x_p)]^n F(\al)\dv\nn
\label{intermed}
\eea
The $n$-summation can be performed using $(1-z)^{-1} =
\sum_{k=0}^\infty  z^{k}$. Finally we perform the $t$-integration
formally using the identity 
\be
\frac{1}{2\pi i}\oint dt\ \frac{e^t}{t-O(\a)} = e^{O(\a)}\ ,
\ee
where $O(\a)$ is an operator depending on $\a$. Here we need to keep
in mind that we are interested in evaluating derivatives with respect
to $\a$ at $\a=0$. This yields the result \r{a1} for $m=0$. 
The induction step goes as follows. We rewrite $z_m$ as
\be
z_m=\nddv \x_p^{m-1} [\x_p,e^{\frac{2}{\pi}\La \x\al} F(\al)]\dv\ ,
\ee
where we used that $\x_p\dv=0$. Evaluating the commutator using
\be
\x_p f(\al)\dv = \cot(2\eta)f^\prime(\al)\dv\ ,
\label{xp}
\ee
where $f^\prime$ is the derivative of $f$ and collecting terms we
obtain
\be
z_m=\frac{1}{\kappa}\nddv \x_p^{m-1}\left[
\cot(2\eta)e^{\frac{2\La}{\pi}\x\al}F^\prime(\al)+\frac{2\La}{\pi}\cot(2\eta)
ye^{\frac{2\La}{\pi}\x\al}F(\al)\right]\dv\ .
\ee
Using the induction assumption and again \r{xp} then yields the
desired result \r{a1}.
Let us now demonstrate how to evaluate the leading contributions to
\r{app2} for $\langle Q_1(m)\rangle$ and $\langle
(Q_1(m))^2\rangle$. They are given by
\bea
\langle (Q_1(m))^l\rangle\bigg|_{\rm lead} &=&
\frac{\partial^l}{\partial\a^l}\bigg|_{\a=0}
\nddv e^{\frac{2}{\pi}\La\x\al}e^{\frac{\al^2}{2\pi^2}\ln(4\La y)}
(g(\n))^2\dv {\left(1+\frac{2\La}{\pi}\cot(2\eta)+{\cal
O}(\La^2)\right)}\ .\nn
\eea
We apply \r{a1} with $m=0$ and $F(\a) =
e^{\frac{\al^2}{2\pi^2}\ln(4\La y)}(g(\n))^2$, then perform the
differentiations with respect to $\a$ and set $\a$ to zero, and
finally use \r{xp} to evaluate the expectation value using the fact
that $\nddv \al_q =0$, $\al_p\dv =0$. The function $g$ has the
properties that $g(0) =1$, $g^\prime(0)=0$ and $g^{\prime\prime}(0)
=2(1+\gamma)$, where $\gamma$ is Euler's constant, which leads to the
result 
\bea
\langle Q_1(m)\rangle\bigg|_{\rm lead} &=&\lam m|\cot\eta|
{\left(1+\frac{6\La}{\pi}\cot(2\eta)+{\cal
O}(\La^2)\right)}\nn
\langle (Q_1(m))^2\rangle\bigg|_{\rm lead} &=&
\left((\lam m|\cot\eta| )^2+\frac{\ln(4\La m|\cot\eta|
)+1+\gamma}{\pi^2}\right)\times\nn
&&\times(1+\frac{8\La\cot(2\eta)}{\pi} +{\cal O}(\La^2)).
\eea
The contributions of the subleading terms can be taken into account in
an analogous way, which leads to the results \r{qs}. We note that
contributions from further subleading terms are of higher order in
$\La$ and $y$.

\newpage

\setlength{\baselineskip}{14pt}

\end{document}